\documentclass[aps, 10pt, prl, twocolumn, amssymb, amsmath, amsfonts, showpacs, superscriptaddress, preprintnumbers, a4paper]{revtex4-1}
\usepackage{amsmath, amssymb, amsfonts, amsthm}
\usepackage{times}
\usepackage{xcolor}
\usepackage{graphicx}
\usepackage{comment}
\usepackage[latin1]{inputenc}
\usepackage[
  breaklinks,
  colorlinks=true,
  bookmarks=true,
  bookmarksopenlevel=3,
  bookmarksnumbered=true,
  menucolor=blue,
  linkcolor=blue,
  citecolor=blue,
  urlcolor=blue,
  pdfcreator={GhostScript},
  pdfkeywords={},
  pdfstartview=FitH]{hyperref}
\usepackage{physics, xparse}

\begin{document}

\def\lptms{Universit\'e Paris-Saclay, CNRS, LPTMS, 91405, Orsay, France.}
\def\lps{Universit\'e Paris-Saclay, CNRS, Laboratoire de Physique des Solides, 91405, Orsay, France.}

\title{Two-fluid coexistence in a spinless fermions chain with pair hopping}

\author{Lorenzo Gotta}\email{lorenzo.gotta@universite-paris-saclay.fr}\affiliation{\lptms}
\author{Leonardo Mazza}\affiliation{\lptms}
\author{Pascal Simon}\affiliation{\lps}
\author{Guillaume Roux}\affiliation{\lptms}

\date{\today}

\begin{abstract}
We show that a simple one-dimensional model of spinless fermions with pair hopping displays a phase in which a Luttinger liquid of paired fermions coexists with a Luttinger liquid of unpaired fermions.
Our results are based on extensive numerical density-matrix renormalisation group calculations
and are supported by a two-fluid model that captures the essence of the coexistence region. 
\end{abstract}


\maketitle

The search for zero-energy Majorana modes, which naturally appear in topological superconducting models~\cite{kitaev_unpaired_2001}, has raised a remarkable interest in the problem of pairing in number-conserving models~\cite{Fidkowski_2011, Sau_2011, Kraus_2013, ortiz_many-body_2014, kells_multiparticle_2015, lang_topological_2015, keselman_gapless_2015, iemini_localized_2015, iemini_majorana_2017, Zhiyuan_2017, Leggett_2017, Leggett_2018, Yin_2019, Lapa_2020}. 
A paired phase is a phase where two (or more) fermions bind together and behave as a singular molecular object. In one dimension (1D), where most of the attention has concentrated so far, the characteristic signature of pairing is the absence of any fermionic order, whereas pairs display quasi-long-range order. For spin-$1/2$ fermions, the attractive Hubbard model naturally favors onsite singlet pairing~\cite{Lee1988,Guerrero2000}. Increasing the number of internal degrees of freedom allows a pairing mode to coexist with a remaining decoupled fermionic mode~\cite{Azaria2009}. For spinless fermions, pairing requires finite-range interaction but no coexistence with unpaired fermions is observed~\cite{mattioli_cluster_2013, dalmonte_cluster_2015, he_emergent_2019, Gotta_2020}. Importantly, spatial interfaces between paired and unpaired phases should host Majorana zero modes, which could then be realised without resorting on superconducting proximity effects~\cite{Ruhman_2015, Ruhman_2017}.
 
The difficulty in studying the pairing transition is that it implies a reshape of the low-energy sector of the model, with the appearance (or disappearance) of Fermi points, to be taken into account by
unconventional bosonisation treatments~\cite{Ruhman_2015, Ruhman_2017}.
A particularly visual model based on two fluids, a bosonic one describing the pairs, and a fermionic one describing the unpaired fermions, has been presented recently~\cite{Kane_2017}.
These studies agree on the fact that paired and unpaired phases are separated by a continuous phase transition with central charge $c=3/2$~\cite{Ruhman_2015, Kane_2017} originating from a standard gapless mode and an additional Ising/Majorana degree of freedom. 
This prediction has been verified by several numerical analyses~\cite{mattioli_cluster_2013, dalmonte_cluster_2015, he_emergent_2019, Gotta_2020}. 

In this Letter, we show that the phenomenology of the pairing transition is richer.
We revisit a 1D spinless-fermion model introduced in Ref.~\cite{Ruhman_2017} in which pair hopping competes with single fermion hopping. 
Related electronic models with correlated hopping, such as the Penson-Kolb-Hubbard model~\cite{Penson1986,Affleck1988,Doniach_1993,Sikkema1995,Japaridze_2001}, have been proposed in  the context of high-$T_c$ superconductors \cite{Hirsch_1990} and lead to rich and complex phase diagrams \cite{Aligia_1997}; our model also bears some relations with the folded spin-1/2 model~\cite{Zadnik_2020, Zadnik_2020b} and the Bariev model, which are exactly-solvable with Bethe ansatz~\cite{Bariev_1991}, and with models for ultra-cold gases with synthetic dimension~\cite{Bilitewski_2016, Chhajlany_2016}.

We show the emergence of a coexistence phase comprising neighbouring paired fermions in a sea of unpaired fermions that is stable towards phase separation. 
Since pairs are composed of two fermions, it is not obvious that they could coexist with gapless fermionic excitations.
Indeed, semiclassical intuition and the standard Luttinger liquid (LL) approach lead to the conclusion that all fermions are either paired or unpaired.
Yet, taking superfluids as a paradigmatic example, phases with two coexisting fluids are not novel to condensed-matter physics~\cite{pitaevskii_2003}.
Our findings are supported by numerical simulations, which are fully interpreted with a phenomenological two-fluid (2F) model inspired by Ref.~\cite{Kane_2017}.
In particular, we clearly pinpoint under which conditions the two kinds of scenarios, extended coexistence phase or a $c=3/2$ transition point, take place.
Such a discovery of the first realisation of a 2F model for describing a 1D phase with a pairing instability opens the path to novel investigations in the context of number-conserving Majorana fermions.

\begin{figure}[t]
\centering
\includegraphics[width=0.95\columnwidth,clip]{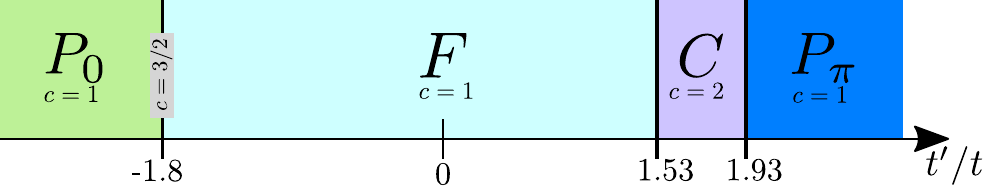}
\caption{Sketch of the phase diagram of model~\eqref{hamiltonian} for density $n=0.25$. Four phases appear: a regular LL fermionic phase $F$, paired LL phases $P_0$ and $P_\pi$ and a coexistence phase $C$ with central charge $c=2$ where fermions and $P_\pi$ pairs are mixed.}
\label{fig:phasediag}
\end{figure}

\paragraph{Hamiltonian.}
We consider a chain of length $L$ with spinless fermion operators $c_j^{(\dagger)}$ and study the model introduced in Ref.~\cite{Ruhman_2017}:
\begin{equation} 
\label{hamiltonian}
H=-t\sum_{j}\left[c^{\dag}_{j}c_{j+1}+\text{h.c.}\right]-t^{\prime}\sum_{j}\left[c^{\dag}_{j+1}c^{\dag}_{j}c_{j}c_{j-1}+\text{h.c.}\right],
\end{equation}
in which $t$ is the fermionic hopping amplitude while $t^{\prime}$ is the pair-hopping amplitude.
The phase diagram only depends on the ratio $\tau = t'/t$ and the density $n =N/L$ with $N$ the total number of fermions. 
The unusual $t^{\prime}$ term favours a gain in kinetic energy for paired configurations, that naturally competes at low densities with the single-fermion kinetic energy term (a similar term has been identified in cold-atoms setups with synthetic dimensions~\cite{Bilitewski_2016, Chhajlany_2016}).
We take $n=0.25$ in the following and analyse such competition with the density-matrix renormalisation-group (DMRG) algorithm~\cite{White1992,White1993,schollwock_density-matrix_2005, schollwock_density-matrix_2011} using two implementations, one of which being the ITensor library~\cite{itensor}.
The obtained phase diagram is sketched \textcolor{blue}{in} Fig.~\ref{fig:phasediag}. 
For small $\tau$ a regular fermionic LL phase $F$ extends from the free fermion point.
At large $|{\tau}|$, two fully paired LL phases $P_0$ and $P_\pi$ are stabilized.
Their main difference is that pairs quasi-condense around either the $k=0$ or the $k=\pi$ momenta.
All these three phases display a central charge $c=1$ corresponding to a single bosonic mode description. For $\tau<0$, it has been shown~\cite{Ruhman_2017} that the transition from $F$ to $P_0$ is direct and features an extra Majorana degree of freedom revealed from the $c=3/2$ central charge.
The main result of this Letter is to show that for $\tau>0$, there is an intervening coexistence phase denoted by $C$ where a LL of $P_\pi$ pairs coexists with a LL of fermions.

\paragraph{Paired phases.}
We first analyse the paired phases exploiting the fact that model~\eqref{hamiltonian} can be diagonalised exactly~\cite{Chhajlany_2016, Zadnik_2020} for $t=0$.
Since the pair-hopping term enhances the kinetic energy of pairs, we assume that the ground state lies in the subspace $\mathcal{H}_{P}$ spanned by states with the $2N_b$ fermions forming $N_b$ nearest-neighbour pairs. Within $\mathcal H_P$, each fermionic state is mapped onto a spin-1/2 configuration over  a lattice of length $L_b =L-N_b$, via the rules $\ket{\bullet\bullet} \rightarrow \ket{\uparrow}$, $\ket{\circ}\rightarrow \ket{\downarrow}$. 
In $L_b$, the $N_b$ term can be understood as an excluded volume.
Then, a spin up stands for a pair while a spin down stands for an empty site.
The action of Hamiltonian~\eqref{hamiltonian} over $\mathcal{H}_{P}$ is unitarily equivalent to that of an effective XX spin-1/2 Hamiltonian $H_{\rm eff} = t^{\prime}\sum_{j=1}^{L_b}\left[\sigma_{j}^{+}\sigma_{j+1}^{-}+\text{h.c.}\right]$. Using Jordan-Wigner transformation and Fourier transform, we readily find  the diagonal form $H_{\rm eff}=\sum_{k} \varepsilon_{p} (k) \,n_{k}$, with the pair band dispersion relation $\varepsilon_{p}(k) = 2 t' \cos(k)$. 
For $t'<0$, the groundstate energy per site $e_{{\rm eff}}=\ev*{H_{\rm eff}}/L$ reads
\begin{equation} \label{XX_GS_energy}
e_{{\rm eff}}=\frac{1}{L}\!\!\!\sum_{|k|< \pi \frac{N_b}{L_b}}\!\!\varepsilon_{p}(k)= -\frac{ 2|t^{\prime}|}{\pi}\left(1-\frac{n}{2}\right)\sin\left( \frac{\pi n}{2-n}\right)
\end{equation} 
where we use the relation $n=2N_b/L$. For $t'>0$, one actually has the same result because the unitary transformation $c_{j}\rightarrow e^{i\frac{\pi}{2}j} c_{j}$ implements the mapping $H(t=0,t^{\prime})\rightarrow H(t=0,-t^{\prime})$.
This result \eqref{XX_GS_energy} is validated by the numerics~\cite{SuppMat}.

\begin{figure}[t]
\centering
\includegraphics[width=0.95\columnwidth,clip]{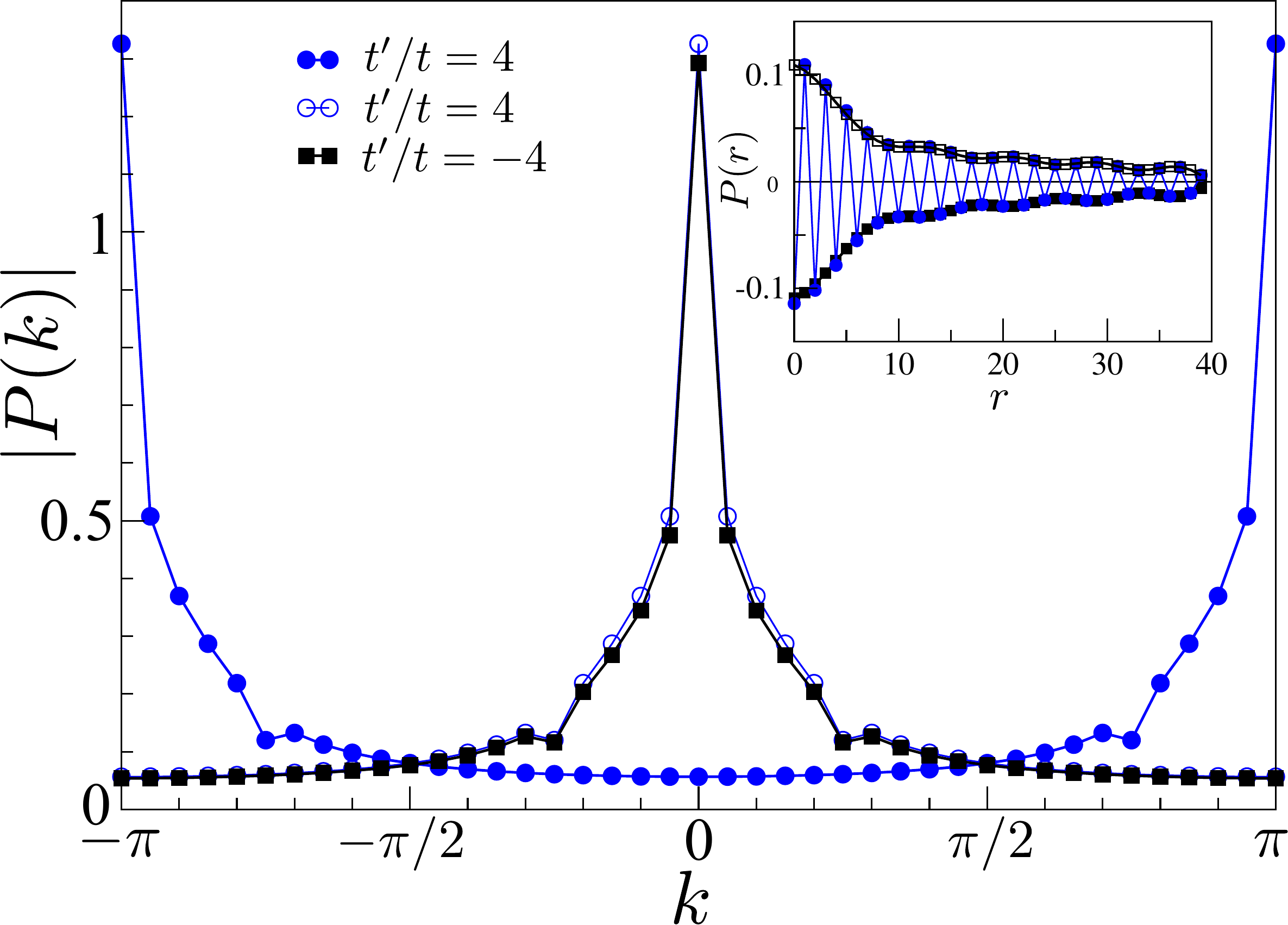}
\caption{Absolute value of the Fourier transform of pair correlations for an open chain with $L=80$ and $t^{\prime}/t=\pm 4$. Open symbols are the $t^{\prime}/t=4$ data shifted by $\pi$.
\textit{Inset:} Pair correlations.}
\label{fig:pair_corr<0}
\end{figure}

The nature of the pairs is qualitatively different in each phase. By inspecting $\varepsilon_p(k)$, we see that the minimum is at $k=\pi$ for $t'>0$, whereas it lies at $k=0$ for $t'<0$. 
The two phases are thus connected by a shift $k \to k+\pi$, corresponding to the application of the  unitary transformation to the pair operator: $c_j c_{j+1} \to  (-1)^j i c_j c_{j+1}$.
This difference persists at finite but large $|\tau|$.
In the inset of Fig.~\ref{fig:pair_corr<0},  we show the pair correlations $P(r) = \ev*{c^{\dag}_{\frac{L}{2}}c^{\dag}_{\frac{L}{2}+1} c_{\frac{L}{2}+r}c_{\frac{L}{2}+r+1} }$ for $\tau = \pm 4$.
They almost exactly coincide in absolute value but differ by a staggering factor $(-1)^r$. 
The main chart of Fig.~\ref{fig:pair_corr<0} displays the pair occupation number $P(k) =\frac{1}{L}\sum_{j,j^{\prime}}e^{ik(j-j^{\prime})}\ev*{c^{\dag}_{j}c^{\dag}_{j+1} c_{j^{\prime}}c_{j^{\prime}+1}}$.
The connection between $P_\pi$ and $P_0$ translates into a shift of the main peak from $k=\pi$ to $k=0$ when changing $\tau=4$ into $\tau=-4$.
Notice that the unitary transformation is no longer valid at non-zero $t$ at the Hamiltonian level. Still, the data show that it becomes an emergent symmetry due to the dominant weights of paired states. Since the transition between the fermionic LL and $P_{0}$ phases has been extensively discussed in Ref.~\cite{Ruhman_2017}, we now focus on $\tau>0$~\footnote{We noticed that the numerical data of Ref.~\cite{Ruhman_2017} actually correspond to the $\tau = t'/t<0$ case.}.

\begin{figure*}
\centering
\includegraphics[width=0.98\textwidth,clip]{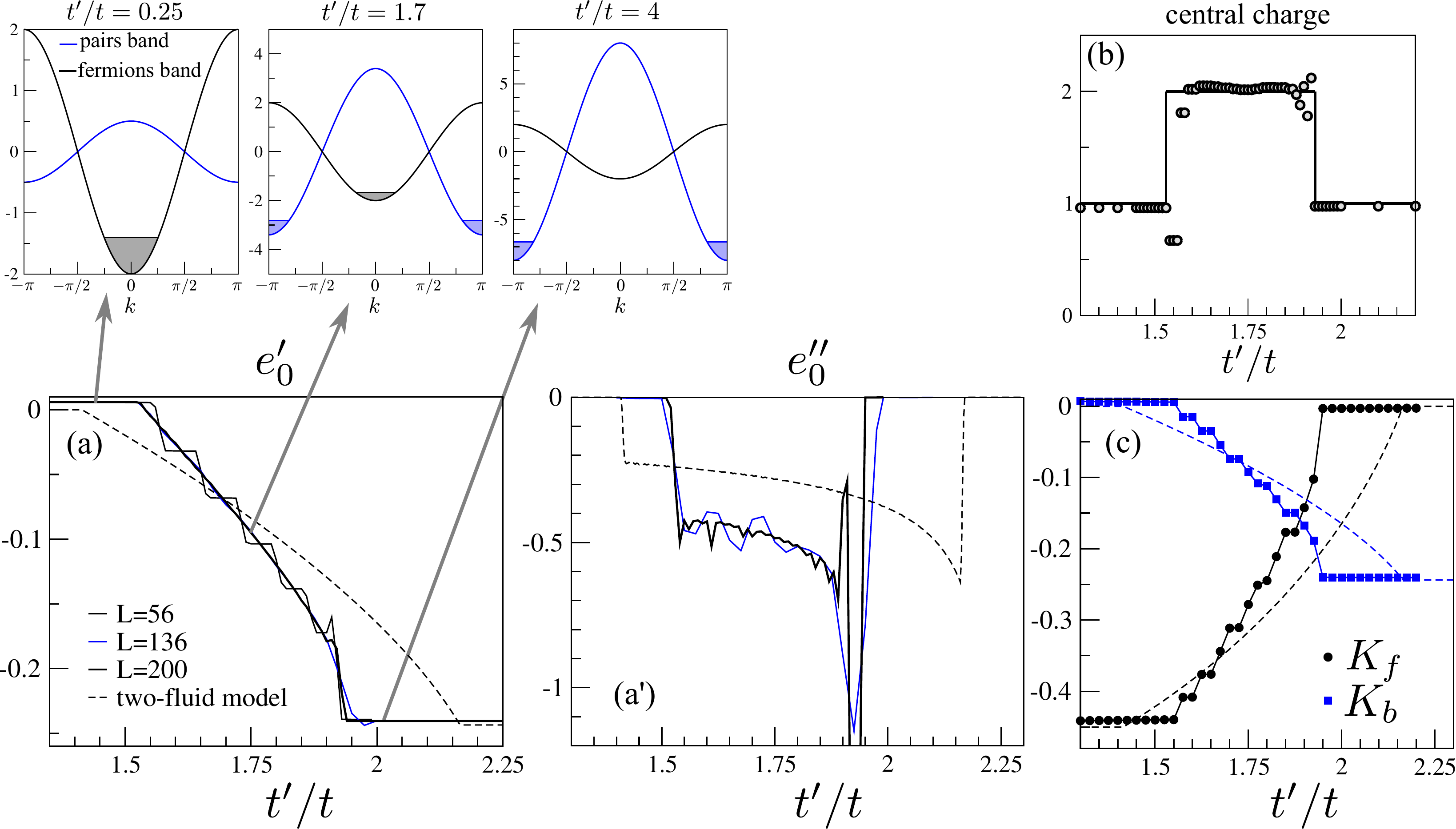}
\caption{(a) First and (a') second derivatives of the energy per site $e_0$ as a function of $\frac{t^{\prime}}{t}$ for three system sizes  $L=56, 136, 200$. Dotted lines are predictions of the 2F model. Arrows point toward typical band structures of the 2F model for $\frac{t^{\prime}}{t}=0.25$, $\frac{t^{\prime}}{t}=1.7$ and $\frac{t^{\prime}}{t}=4$. (b)
Fitted central charges as a function of $\frac{t^{\prime}}{t}$. 
(c) Single-particle kinetic energy $K_f$ and pair kinetic energy $K_b$ probing almost directly $n_f$ and $n_b$.}
\label{fig:c}
\end{figure*}

\paragraph{Coexistence phase.}
We  now present numerical results for the intervening coexistence phase $C$ between the $F$ and $P_\pi$ LL. 
In Fig.~\ref{fig:c}(a) and (a'), we plot the first and second derivatives of the ground state energy per site $e_0(\tau)$. While constant behaviors are found in $F$ and $P_\pi$, a finite intermediate region emerges between two finite jumps of the second derivative. 
Since the first derivative is continuous up to finite-size effects, we observe two continuous phase transitions that mark the existence of the C phase, in contrast with the first order transition scenario proposed in Ref.~\cite{Ruhman_2017} and as will be clear in the following.
Within the grid precision, the boundary of this $C$ phase are found at $\tau_{c1}\simeq 1.53(1)$ and $\tau_{c2}\simeq 1.93(1)$.
Last, increasing the system sizes shows that the phase is stable and does not shrink as can be seen in the figure.

A first insight in the nature of the $C$ phase is presented in Fig.~\ref{fig:c}(b).
We show that the central charge $c$, estimated from fits of the entanglement entropy~\cite{SuppMat}, jumps from $c=1$ in $F$ and $P_{\pi}$ to the value $c=2$ in the $C$ phase.
These values indicate that in the $F$ and $P_{\pi}$ phases have a single effective bosonic mode, whereas the $C$ phase possesses 2 bosonic modes that will be identified in the following.
In the rest of the article, we develop an effective model that (i) captures the low-energy physics of Hamiltonian~\eqref{hamiltonian}, (ii) explains the nature of the $C$ phase, and (iii) elucidates the different mechanisms at play between the $\tau<0$ and $\tau>0$ branches of the phase diagram.

\paragraph{Two-fluid model.} 
We start by assuming that the system is composed of 2 species of particles, one fermionic (the unpaired fermions) and one bosonic (the pairs), described respectively by a free fermion Hamiltonian $H_{f}$ and an XX model $H_{b}$:
\begin{align}
 H_{f} =& -t \sum_j d_j^\dagger d_{j+1}+ \text{h.c.} \;, \\
 H_{b} =& + t' \sum_j \sigma^+_j \sigma^-_{j+1} + \text{h.c.} \;.
\end{align}
It is important to stress that this is an effective model, and that the $d_j$ fermions (satisfying canonical anticommutation relations) do not coincide with the original ones because they only describe the unpaired particles.
This assumption is motivated by the limiting properties of Hamiltonian~\eqref{hamiltonian} for $t=0$ and $t'=0$ that we discussed above. We stress however that there is no exact handy mapping onto~\eqref{hamiltonian}: the two-fluid (2F) model $H_{2F} = H_{f}+H_{b}$ has a phenomenological nature.

As a minimal model, we further assume that the two species interact only through the total density constraint $n=n_{f}+2n_{b}$, where $n_{f,b}=N_{f,b}/L$ are the effective fermionic and bosonic densities. The ground state energy per site $e_{2F}$ is then the sum of the fermionic and bosonic contributions:
\begin{equation}
e_{2F}=-\frac{2t}{\pi}\left[\sin\left(\pi n_{f}\right)+\tau \sin\left(\pi\frac{n-n_{f}}{2}\right)\right].
\label{Eq:2F:GS}
\end{equation}
By minimizing $e_{2F}$ with respect to the free parameter $n_f$ using standard techniques~\cite{SuppMat}, we identify three regions that are depicted in the sketches of Fig.~\ref{fig:c}:
(i) a fully fermionic region, for $0< \tau< \tau_{c1}$ with $\tau_{c1} = 2 \cos(\pi n) \simeq 1.41$, in which $n_f = n$ and $n_b=0$, that we associate to the $F$ phase;
(ii) an intermediate region, for $\tau_{c1} < \tau < \tau_{c2}$ with $\tau_{c2} =2 / \cos(\pi n/2) \simeq 2.16$, in which both $n_f$ and $n_b$ are non-zero, that we associate with the $C$ phase;
(iii) a fully bosonic region, for $\tau > \tau_{c2} $, in whic $n_f=0$ while $n_b = n/2$, corresponding to the $P_\pi$ phase.
Incidently, the natural order parameter through the phase diagram is $n_f$ -- or equivalently $n_b$.
In this two-fluid picture, (ii) is naturally a region of \textit{coexistence} of the bosonic and fermionic fluids, hence the name.
 
The 2F model thus proposes an interpretation of the two transition points in terms of two band-filling (band-emptying) Lifshitz transitions, which are associated to the appearance (disappearance) of two Fermi points as sketched in Fig.~\ref{fig:c}(a).
We are thus in front of two continuous and second-order quantum phase transitions, and not of a first-order transition, which is another scenario that would be \textit{a priori} possible.
In the following, we show that the two-fluid model provides a good description of the $C$ phase.

\paragraph{Interpretation of numerical data.}
We first observe that the 2F model describes in a natural way the DMRG data of Fig.~\ref{fig:c}. The main difference is that the boundary points are not quantitatively reproduced.
Let us start with the central charge: in the 2F model, the coexistence phase has $c=2$ that corresponds to two effective low-energy bosonic fields in LL theory.
These modes stem from the effective existence of two gapless Fermi points $k_f$, leading to a single bosonic mode in LL theory, and two "hard-core boson" Fermi points $k_b$ adding up another bosonic mode. On the contrary, the $F$ and $P_{\pi}$ region only have two effective Fermi points, leading to standard $c=1$ phases. This agrees perfectly with the numerical data in Fig.~\ref{fig:c}(b).

\begin{figure}[t]
\centering
\includegraphics[width=\columnwidth,clip]{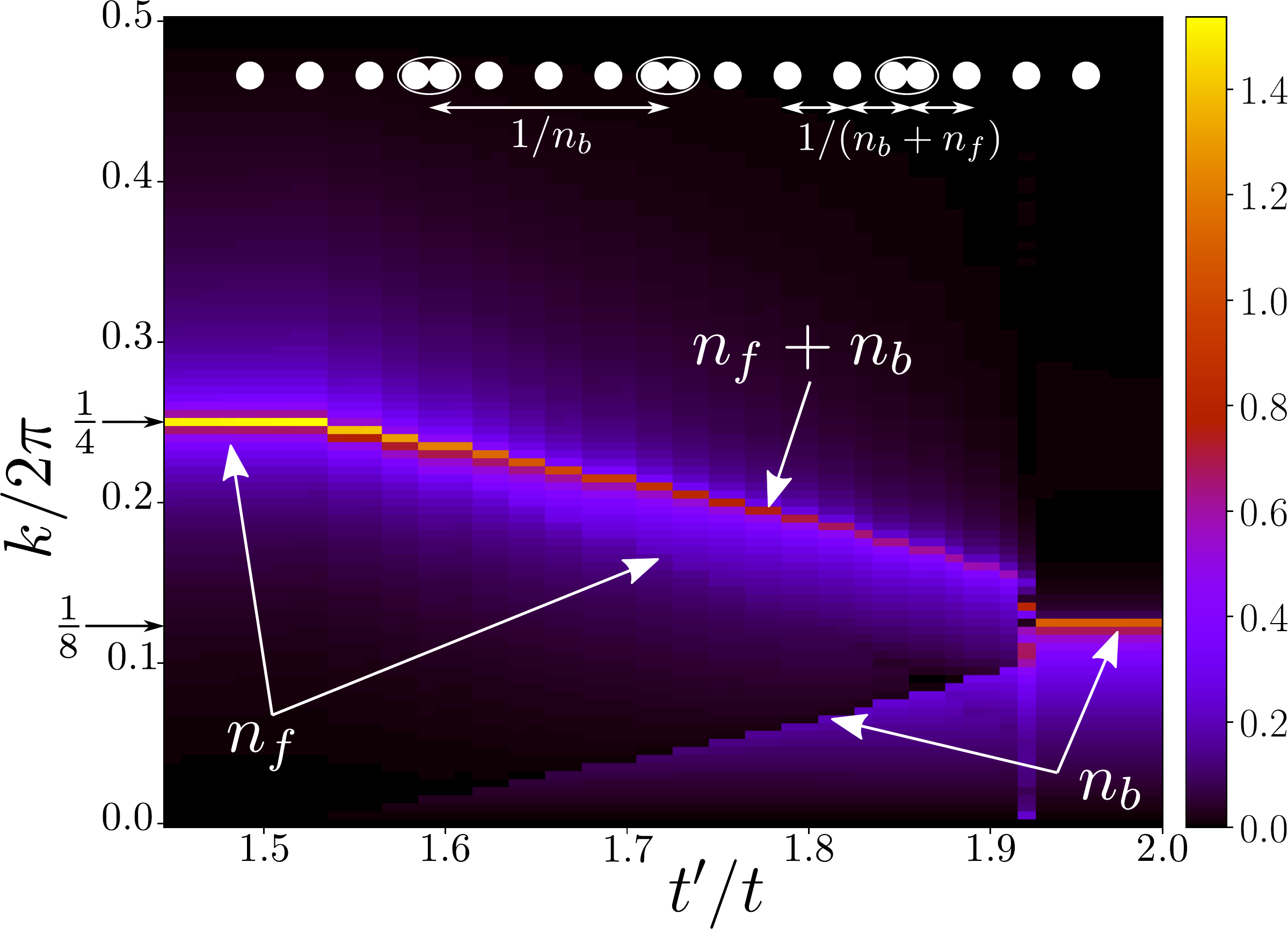}
\caption{Map of the absolute value of the Fourier transform of local density fluctuations $\langle\delta n_{j}\rangle$ as a function of $t^{\prime}/t$ for an open chain with $L=200$. }
\label{fig:dn_k}
\end{figure}

We now focus on the comparison with local observables to further characterize the $C$ phase. 
Focusing on the energy, we superimpose the 2F model prediction for the first and second derivatives $e_{2F}'(\tau)$ and $e_{2F}''(\tau)$ to the DMRG data in Fig.~\ref{fig:c}(a) and (a').
We observe two jump discontinuities that are computed exactly in the 2F model~\cite{SuppMat}. 
The qualitative resemblance with the numerical data is impressive, given the simplicity and phenomenological nature of the 2F model.
Furthermore, this total energy splits into two contributions that are direcly connected to the order parameters $n_f$ and $n_b$.
We define the single-particle kinetic energy $K_f = -\frac{1}{L} \sum_j \ev*{c_j^\dagger c_{j+1}+\text{h.c.}}$ and the pair hopping kinetic energy $K_b = -\frac{1}{L}\sum_j\ev*{c_{j+1}^\dagger c_{j}^\dagger c_j c_{j-1}+\text{h.c.}}$ such that $e_0 = tK_f + t'K_b$ and $K_b = e_0'(\tau)$ according to Feynman-Hellmann theorem. The 2F model prediction then simply corresponds to each term of Eq.~\eqref{Eq:2F:GS}. The comparison with DMRG data is displayed on Fig.~\ref{fig:c}(c) with dot lines. $K_f$ and $K_b$ capture the order parameters value essentially up to a sine function. We do observe that they are very close to zero in the $P_{\pi}$ and $F$ phases, respectively. 
In the $C$ phase, they are both varying following the qualitative behaviour  obtained within the 2F model. Lastly, the band filling interpretation helps understand finite-size effects: in both Fig.~\ref{fig:c}(a) and (c), jumping from one plateau to the next corresponds to filling the system with another pair. For instance with $L=56$, there are between 0 to 7 pairs that are progressively created as $t'$ increases.

This pair creation is also well seen in the density profile obtained with open boundary conditions with DMRG, and which fully supports our interpretation of the $C$ phase.
We show in Fig.~\ref{fig:dn_k} the Fourier transform of the local density fluctuations $\ev*{\delta n_{j}} = \ev*{c^{\dag}_{j}c_{j}}-n$.
Indeed, for the $F$ and $P_\pi$, we expect leading fluctuations at $2k_f = 2\pi n$ and $2k_b = \pi n$ respectively. Such constant behaviors are clearly observed in Fig.~\ref{fig:dn_k} around the coexistence region by recalling that $n=0.25$. 
Within the $C$ phase, the leading fluctuations emerge at $k = 2\pi (n_f + n_b)$, with a second main peak at $k = 2\pi n_b$ and a tiny signal at $k = 2\pi n_f$.
This is understood from the sketch of Fig.~\ref{fig:dn_k}. In the $C$ phase, pairs effectively repel each other and add some excluded volume to the remaining unpaired fermions. Then, if none equally spaces the total number $N_f+N_b$ of effective particles, this corresponds to mean distance of $1/(n_f + n_b)$. On top of that, pairs add an extra signal -- since that locally double the density -- corresponding to a typical spacing of $1/n_b$. Consequently, following the peaks location allows one to quantitatively follow the order parameters. Notice that such excluded volume effects go beyond the 2F model picture according to which the fluctuations of the two fluids should be independent.

\paragraph{Phase stability.}
If we consider that interactions between fermions and bosons have been totally neglected,
the effectiveness of the 2F model looks rather surprising. 
In reality, as we have seen, these degrees of freedom delocalise on the same 1D setup and effectively repel each other because one site cannot be occupied by one fermion and one boson at the same time.
The main consequence of this is that single-particle hopping can create one pair by putting two unpaired fermions close by (and \textit{viceversa}).
At a first level of approximation, we need to include a term like $- t \sum_j (\sigma_j^+ d_j d_{j+1} + \text{h.c.})$.
Yet, one such term is completely irrelevant because it conserves momentum: the annihilation of two fermions with momentum $k_1$ and $k_2$ leads to the creation of one boson with momentum $k_1+k_2+ 2\pi m$, where $m \in \mathbb Z$. As we have discussed initially, the fermionic particles are concentrated around $k \sim 0$,  whereas bosonic ones are located around $k\sim \pi$. 
This term is thus ineffective, it fails to hybridize bosonic and fermionic degrees of freedom, and can be safely neglected.
On the contrary, the bosons quasi-condense around $k=0$ in between the $F$ and $P_0$  phases. Interactions are then resonant and hybridize fermionic and bosonic degrees of freedom. According to the description developed in Ref.~\cite{Kane_2017}, one then expects a direct continuous transition with central charge $c=3/2$, in agreement with the numerics for $t'<0$~\cite{Ruhman_2017}.

\paragraph{Conclusions.}  
We have presented a study of the pairing transition in a model featuring a competition between the delocalisation of fermions and of pairs.
The DMRG results and their interpretation using a simple phenomenological model
strongly support the existence of an unexpected coexistence phase of paired and unpaired fermions.
These remarkable outcomes put on a more solid basis the 2F model presented in Ref.~\cite{Kane_2017} and opens the route to a wider applications in the context of one-dimensional models featuring paired phases.

\paragraph{Acknowledgements.}
We gratefully acknowledge discussions with S.~Capponi, M.~Fagotti, E.~Orignac, J.~Ruhman and L.~Zadnik.
We acknowledge funding by the Agence Nationale de la Recherche (ANR) under the project TRYAQS (ANR-16-CE30-0026) and by LabEx PALM (ANR-10-LABX-0039-PALM). This work has been supported by Region Ile-de-France in the framework of the DIM Sirteq.
 
\bibliographystyle{apsrev4-1}
\bibliography{Twofluid.bib}

\onecolumngrid
\appendix
\newpage

\section*{Supplementary Material}

\subsection{Density in the 2F model}

\begin{figure}[b]
\centering
\includegraphics[width=0.5\columnwidth]{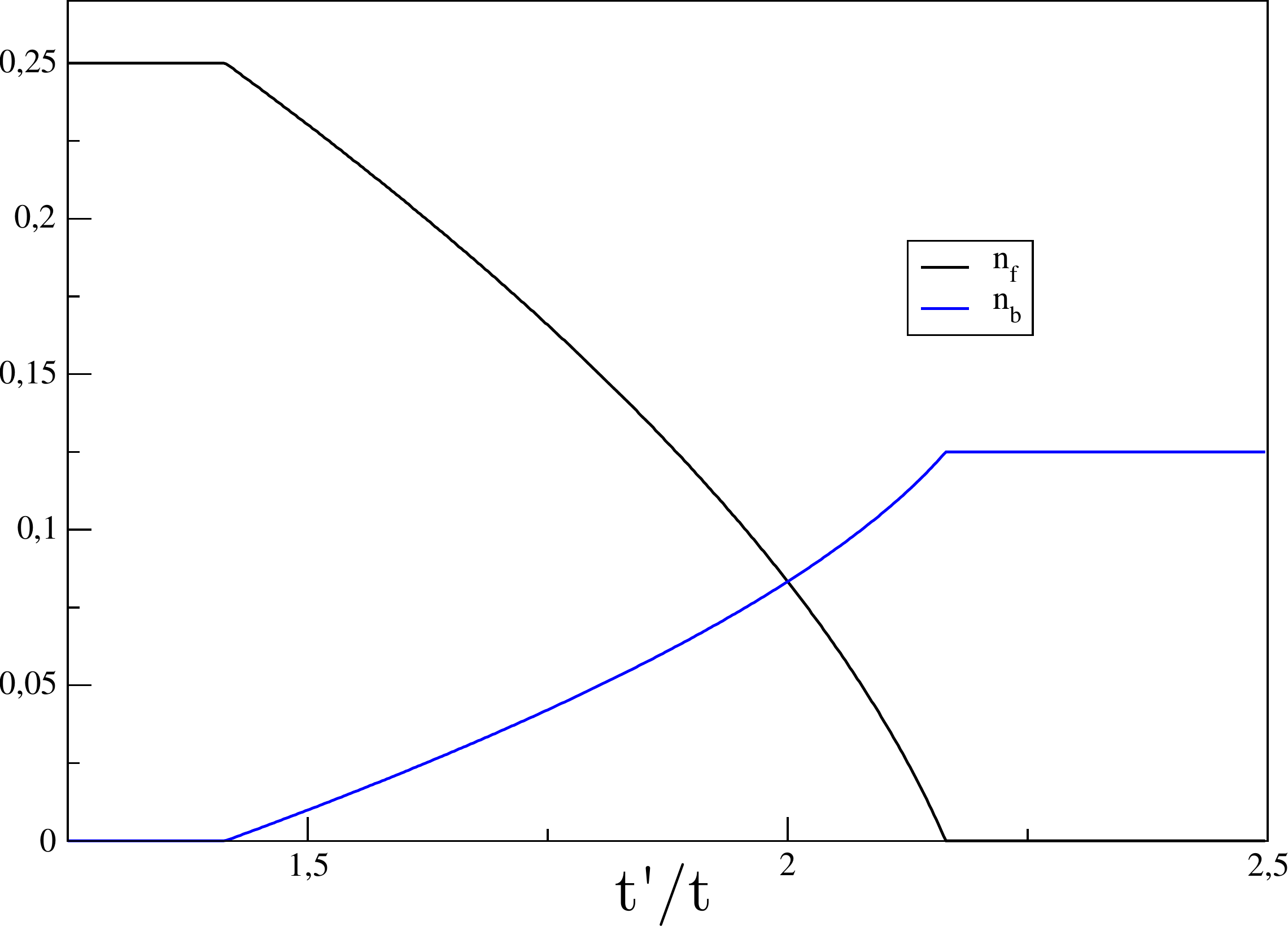}
\caption{Optimal fermionic and bosonic densities obtained by minimizing $\frac{e_{2F}}{t}$ for a total density $n=0.25$.}
\label{fig:supmat_fig1}
\end{figure}

Let us explicitly derive the results presented in the main text on the effective 2F model. The behaviour of the optimal fermionic density can be captured by estimating numerically the value of $n_{f}$ that minimizes the rescaled ground state energy density $\frac{e_{2F}}{t}$, $e_{2F}$ being given in Eq.~\eqref{Eq:2F:GS}, as a function of $\tau =t'/t$. The result is presented in Fig.~\ref{fig:supmat_fig1}, where the purely fermionic region ($n_{f}=n$), the purely bosonic region ($n_{b}=0$) and the mixed region ($0<n_{f}<n$) are clearly depicted.

In order to obtain the behaviour of the optimal fermionic density as approaching the phase boundaries of the mixed phase, we consider the stationarity condition $\displaystyle \dv{e_{2F}}{n_{f}}=0$, which reads:
\begin{equation} \label{stationarity_condition}
\cos(\pi n_{f})=\frac{\tau}{2}\cos\left(\pi\frac{n- n_{f}}{2}\right)
\end{equation}
Assuming self-consistently that $n_{f}\approx n$, which amounts to enforce the system to approach the fully fermionic region from the coexistence phase, the r.h.s. of Eq. ($\ref{stationarity_condition}$) takes the form $\frac{\tau}{2}[1+O\left((n-n_{f})^{2}\right)]$, whereas the l.h.s. reads:
\begin{equation}
\cos(\pi n_{f})=\cos(\pi n)\left[1+O\left((n-n_{f})^{2}\right)\right]-\sin(\pi n)\left[\pi (n_{f}-n)+O\left((n-n_{f})^{2}\right)\right].
\end{equation}
Neglecting the terms proportional to $(n_{f}-n)^{m}$ for $m>1$ and solving for the fermionic density $n_{f}$, one obtains:
\begin{equation}\label{density_asymptotics1}
n_{f}(\tau)\approx n-\frac{\frac{\tau}{2}-\cos(\pi n)}{\pi \sin(\pi n)},
\end{equation}
which allows to identify the location of the critical point separating the fermionic phase and the mixed phase with 
\begin{equation}
\tau_{c1}=2\cos(\pi n) \simeq 1.41
\end{equation}
and justifies neglecting higher order contributions in $n_{f}-n$ as they would scale as increasing powers in the deviation from the critical point $\tau-\tau_{c1}$.

Similarly, assuming $n_{f}\approx 0$, we can explore the asymptotic behaviour of $n_{f}$ as the system approaches the transition to the fully bosonic phase. In this case, the l.h.s. of Eq. ($\ref{stationarity_condition}$) reads $1+O\left((n_{f})^{2}\right)$, while the r.h.s. takes the form:
\begin{equation}
\frac{\tau}{2}\cos\left[\pi\frac{n-n_{f}}{2}\right]=\frac{\tau}{2}\left\{\cos\left(\frac{\pi n}{2}\right)\left[1+O(n_{f}^{2})\right]+\sin\left(\frac{\pi n}{2}\right)\left[\frac{\pi n_{f}}{2}+O(n_{f}^{2})\right] \right\}.
\end{equation}
Neglecting terms of order higher than one in the density, we derive the asymptotic behaviour:
\begin{equation}\label{density_asymptotics2}
n_{f}(\tau)\approx\frac{1-\cos(\frac{\pi n}{2})\frac{\tau}{2}}{\frac{\pi}{2}\sin(\frac{\pi n}{2})\frac{\tau}{2}},
\end{equation}
from which we extract the value of the critical point 
\begin{equation}
\tau_{c2}=\frac{2}{\cos\left(\frac{\pi n}{2}\right)} \simeq 2.16
\end{equation}
separating the mixed phase from the bosonic phase and we justify a posteriori the truncation of the Taylor expansion in powers of the density.

\subsection{Energy in the 2F model}

Let us consider the rescaled energy density in the 2F model:
\begin{equation}
\epsilon_{GS}\left(\tau,n_{f}(\tau)\right)=\frac{e_{2F}}{t}=-\frac{2}{\pi}\left[\sin(\pi n_{f}(\tau))+\tau\sin\left(\pi\frac{n-n_{f}(\tau)}{2}\right)\right],
\end{equation}
where $n_{f}(\tau)$ is the optimal fermionic density as a function of $\tau$. Its first derivative with respect to $\tau$ can be written as:
\begin{equation}
\dv{\epsilon_{GS}\left(\tau,n_{f}(\tau)\right)}{\tau}=\frac{\partial \epsilon_{GS}\left(\tau,n_{f}(\tau)\right)}{\partial \tau}+\frac{\partial \epsilon_{GS}\left(\tau,n_{f}\right)}{\partial n_{f}} \bigg\rvert_{n_{f}=n_{f}(\tau)}\dv{n_{f}(\tau)}{\tau}=\frac{\partial \epsilon_{GS}\left(\tau,n_{f}(\tau)\right)}{\partial \tau}=-\frac{2}{\pi}\sin\left(\pi \frac{n-n_{f}(\tau)}{2}\right),
\end{equation}
where the stationarity condition $\displaystyle \frac{\partial\epsilon_{GS}\left(\tau,n_{f}\right) }{\partial n_{f}}\bigg\rvert_{n_{f}=n_{f}(\tau)}=0$ satisfied by the optimal fermionic density has been used to get rid of the second contribution.

\begin{figure}
\centering
\includegraphics[width=0.5\columnwidth]{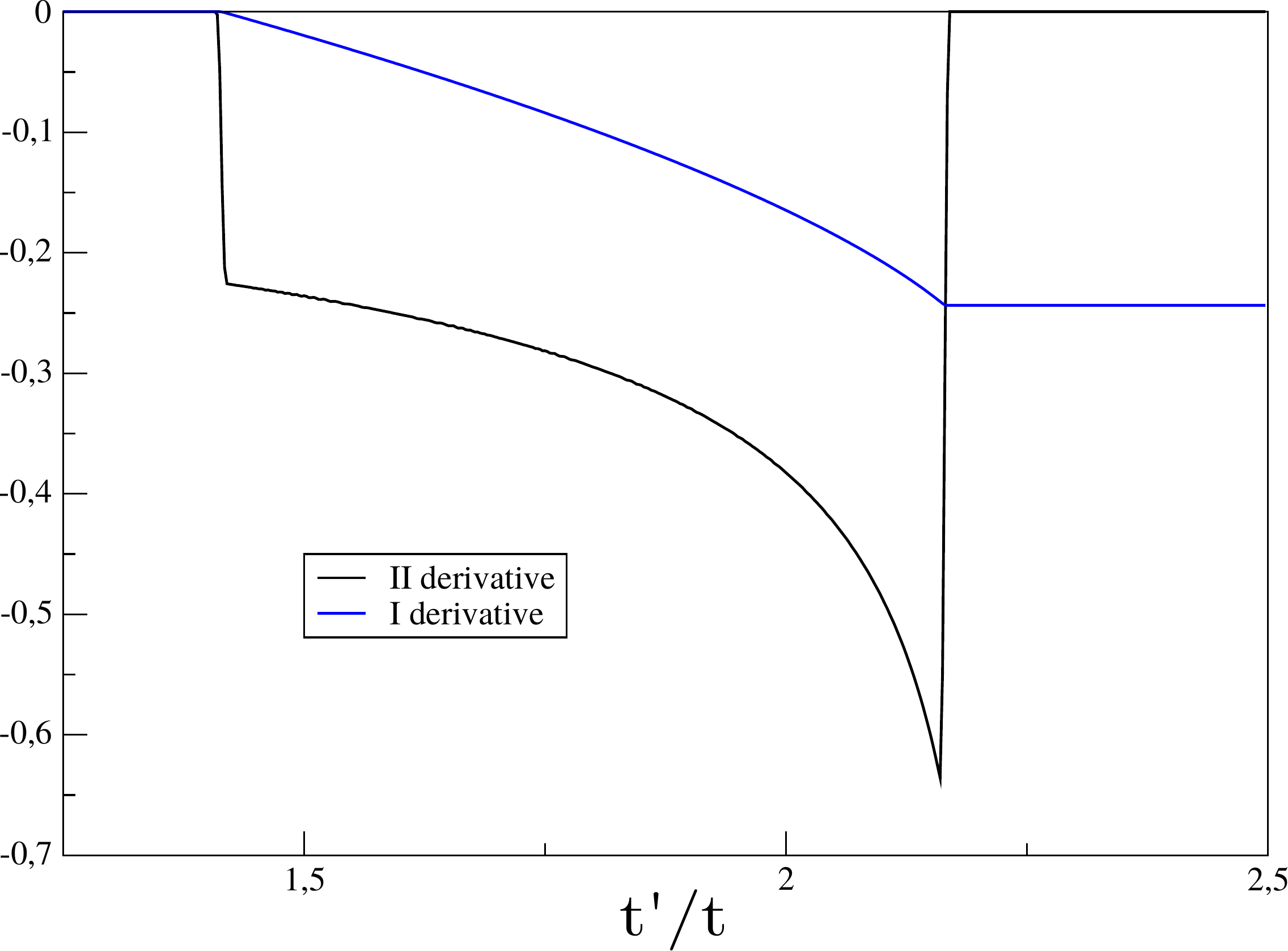}
\caption{First and second derivative of $\epsilon_{GS}\left(\tau,n_{f}(\tau)\right)$ as a function of $\tau$ for a total density $n=0.25$.}
\label{fig:supmat_fig2}
\end{figure}

Consequently, given that $n_{f}=n$ in the fermionic phase, $n_{f}=0$ in the bosonic phase and the asymptotic behaviour of $n_{f}$ while approaching the critical point from the mixed phase is given in Eqs. ($\ref{density_asymptotics1},\ref{density_asymptotics2}$), it is straightforward to obtain that:
\begin{equation}
\dv{\epsilon_{GS}\left(\tau,n_{f}(\tau)\right)}{\tau}= \begin{cases}
0 & 0\leq \tau\leq 2\cos(\pi n) \\
-\frac{1}{2\pi\sin(\pi n)}\left(\tau-2\cos(\pi n)\right) &  \tau\rightarrow \left(2\cos(\pi n)\right)^{+} \\
-\frac{2}{\pi}\left[\sin\left(\frac{\pi n}{2}\right)+\frac{\cos^{3}\left(\frac{\pi n}{2}\right)}{2\sin\left(\frac{\pi n}{2}\right)}\left(\tau-\frac{2}{\cos\left(\frac{\pi n}{2}\right)}\right)\right] & \tau\rightarrow \left(\frac{2}{\cos\left(\frac{\pi n}{2}\right)}\right)^{-} \\
-\frac{2}{\pi}\sin\left(\frac{\pi n}{2}\right) & \tau>\frac{2}{\cos\left(\frac{\pi n}{2}\right)}.
\end{cases}
\end{equation}
From the above formula, the singular behavior of the second derivative of the rescaled ground state energy density can be characterized as follows:
\begin{equation}
\dv[2]{\epsilon_{GS}\left(\tau,n_{f}(\tau)\right)}{\tau}= \begin{cases}
0 & 0\leq \tau\leq 2\cos(\pi n) \\
-\frac{1}{2\pi\sin(\pi n)} &  \tau\rightarrow \left(2\cos(\pi n)\right)^{+} \\
-\frac{\cos^{3}\left(\frac{\pi n}{2}\right)}{\pi\sin\left(\frac{\pi n}{2}\right)} & \tau\rightarrow \left(\frac{2}{\cos\left(\frac{\pi n}{2}\right)}\right)^{-} \\
0 & \tau>\frac{2}{\cos\left(\frac{\pi n}{2}\right)},
\end{cases}
\end{equation}
thus proving that the two transitions are of second order type, as the first derivative is continuous and the second derivative exhibits a finite jump discontinuity at the critical points. The results are summarized in Fig. $\ref{fig:supmat_fig2}$, where the first and second derivative of the ground state energy density are shown as a function of $\tau$.

Last we notice that the jumps in the second derivative of the energy are directly related to jumps in the first derivative of the order parameter $n_f$ since we can show that
\begin{equation}
\dv[2]{\epsilon_{GS}\left(\tau,n_{f}(\tau)\right)}{\tau} = 
\frac{\pi}{2}\cos\left(\frac{\pi}{2}(n-n_f(\tau))\right)\dv{n_f}{\tau}\;.
\end{equation}

\subsection{Estimate for $\tau_{c1}$ taking into account excluded volume}

Extending the 2F model by including the excluded volume exerted by pairs on the unpaired fermions, one can actually get a pretty good estimate for $\tau_{c1}$.  Considering that unpaired fermions have an available volume of $L-2N_b$, one arrives at the energy
\begin{equation}
e_{2F}=-\frac{2t}{\pi}\left[(1-(n-n_f))\sin\left(\pi \frac{n_{f}}{(1-(n-n_f))}\right)+\tau \sin\left(\pi\frac{n-n_{f}}{2}\right)\right].
\label{Eq:2Fextended}
\end{equation}
in which we implicitly assume that $n_b \ll n_f$ so that the energy of pairs remains in the diluted limit since we will take $n_b\to 0$. Then, we obtain
\begin{equation}
\tau_{c1}=2(1-n)\cos(\pi n) + \frac{2}{\pi}\sin(\pi n) \simeq 1.51
\end{equation}
that is very close to the numerical observed value of $\tau_{c1} \simeq 1.53(1)$.

\subsection{Effective XX model}

\begin{figure}
\centering
\includegraphics[width=0.5\columnwidth]{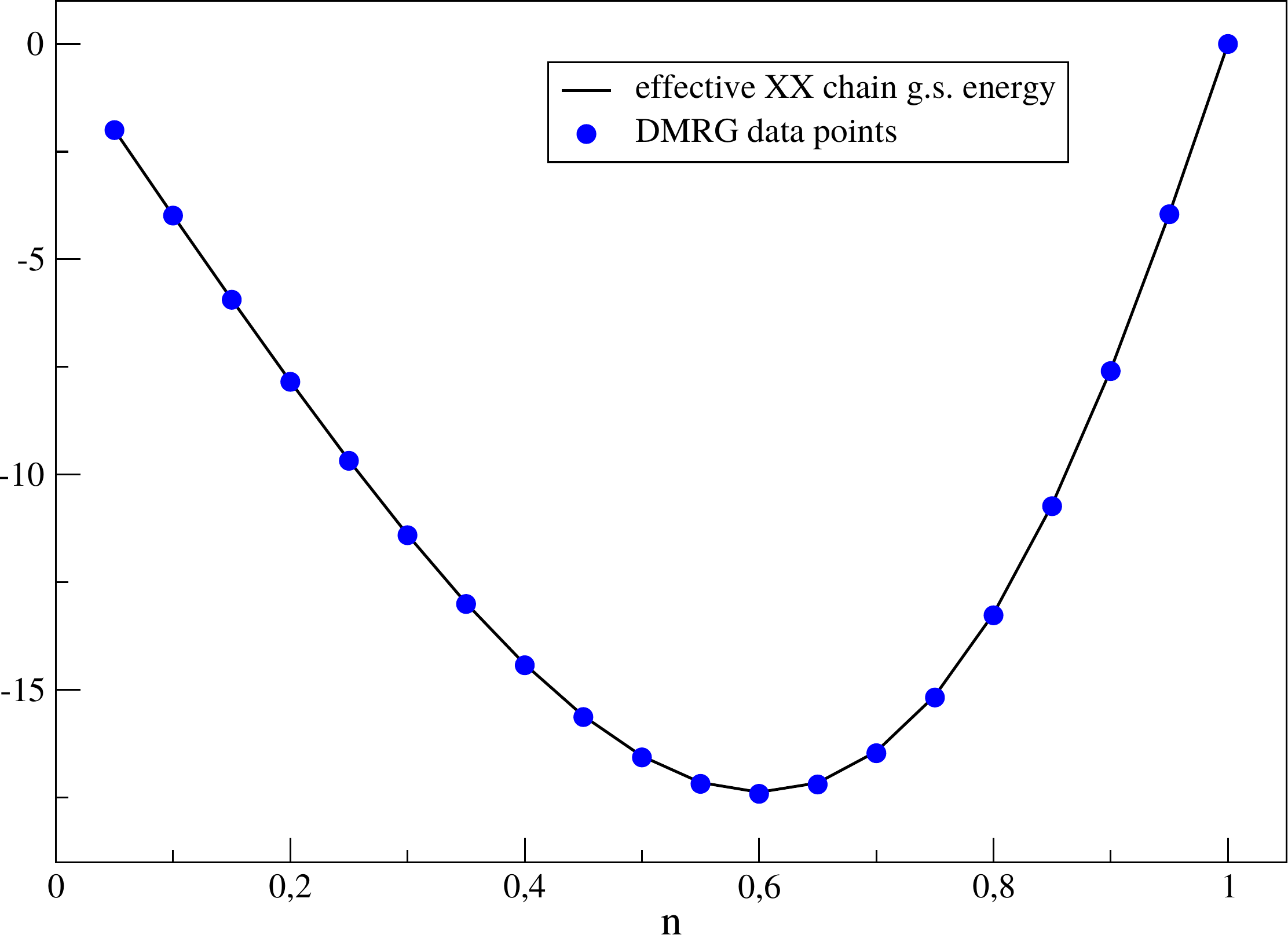}
\caption{Ground state energy  of a system governed by $H_{t^{\prime}}$ with PBC on a lattice of size $L=40$ as a function of the filling $n\in (0,1]$.}
\label{fig:supmat_fig3}
\end{figure}

We test the reliability of the effective XX chain description of the model Hamiltonian in the $t=0$ limit by comparing the ground state energy computed for the Hamiltonian $H_{t^{\prime}}=-t^{\prime}\sum_{j}\left[c^{\dag}_{j+1}c^{\dag}_{j}c_{j}c_{j-1}+h.c.\right]$ via DMRG simulations with the analytical result presented in Eq. ($\ref{XX_GS_energy}$) for the effective XX spin model. In figure ($\ref{fig:supmat_fig3}$), the ground state energy for a system described by the model Hamiltonian $H_{t^{\prime}}$ is successfully fitted by its theoretical estimate in Eq. ($\ref{XX_GS_energy}$) as the filling of the system is varied, certifying as a result the success of the effective spin model in capturing the energetic behaviour of the system for $t=0$.

\begin{figure}[t]
\centering
\includegraphics[width=0.65\columnwidth,clip]{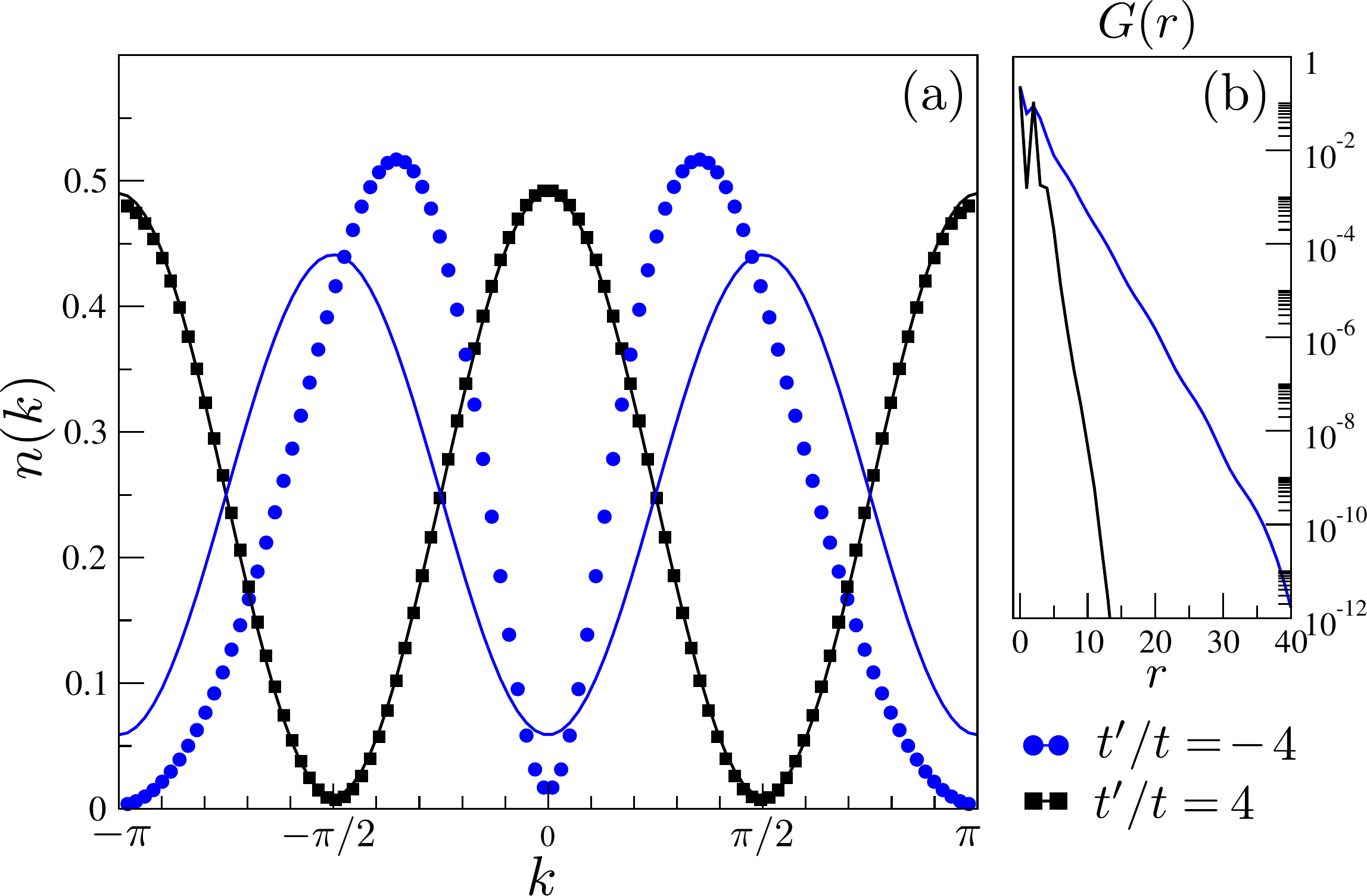}
\caption{Absolute value of the fermionic occupation factor (blue dots) compared to its analytical estimate (orange line). The numerical parameters are $L=80,\,t=1,\,t^{\prime}=4$ with OBC.}
\label{fig:occ_f>0}
\end{figure}
\subsection{Occupation factor}
An additional characterisation of the paired phases comes from the analysis of the fermionic occupation number $n(k)=\frac{1}{L}\sum_{j,j^{\prime}}e^{ik(j-j^{\prime})}\langle c^{\dag}_{j}c_{j^{\prime}}\rangle$. 
It is easy to analytically estimate $n(k)$ for $t=0$ assuming that all fermions are paired: in this case, non-trivial correlations exist only at distance $\pm 2$:
\begin{equation}
\label{Eq:CorrF}
\langle c^{\dag}_{j\prime}c_{j}\rangle = \langle c^{\dag}_{j}c_{j}\rangle\delta_{j,j^{\prime}}+\langle c^{\dag}_{j+2}c_{j}\rangle\delta_{j+2,j^{\prime}}+\langle c^{\dag}_{j-2}c_{j}\rangle\delta_{j-2,j^{\prime}}.
\end{equation}
and originate from the delocalisation of a single tightly-bound pair. Using Eq.~\eqref{Eq:CorrF} we obtain:
\begin{equation} \label{occ_factor_estimate}
n(k)=n+\frac{2}{L} \bigg|\sum_{j}\langle c^{\dag}_{j} c_{j+2}\rangle\bigg|\cos{(2k-\phi)},
\end{equation} 
where $\phi$ is the phase of the complex number $\sum_{j}\langle c^{\dag}_{j} c_{j+2}\rangle$, which in our case is actually real, leading to the sole possibilities $\phi=0,\pi$. Thus, for $\tau = \pm \infty$ the two results differ by a phase $\phi=\pi$. In this limit, the value of $\sum_{j}\langle c^{\dag}_{j} c_{j+2}\rangle$ can be computed as $-\sum_{j}\langle\sigma_{j}^{+}\sigma_{j+1}^{-}\rangle$, where the sum extends over the sites of the effective XX chain lattice of size $L^{\prime}=L-N_{b}$, $N_{b}$ being the number of pairs. The latter expectation value can then be easily computed on the ground state of the XX chain and it shows full consistency with the numerical results for the occupation factor in the $\tau=\pm\infty$ limit of Hamiltonian ($\ref{hamiltonian}$).

On the other hand, the numerical results are compared with the analytical predictions in the strong pairing regime for $\tau= \pm 4$ (where no explicit expression for $\sum_{j}\langle c^{\dag}_{j} c_{j+2}\rangle$ is available) in Fig.~\ref{fig:occ_f>0}, left. No simple qualitative relation is possible between the two cases. Whereas for $\tau=4$ the momentum distribution function resembles the case $\tau=+\infty$, as it is accurately reproduced by Eq. ($\ref{occ_factor_estimate}$) with $\phi=0$ and the numerical value of $\sum_{j}\langle c^{\dag}_{j} c_{j+2}\rangle$, for $\tau = -4$ we obtain data that are radically different from the case $\tau = -\infty$ and are not captured by Eq. ($\ref{occ_factor_estimate}$). We interpret this behaviour as a different rigidity of the two paired phases with respect to kinetic single-particle perturbations: for $\tau>0$ the system is extremely stable, for $\tau<0$ it is more prone to modifications.


It is natural to interpret the nature of these modifications in terms of an increased spatial extension of the pairs of the system. We support this statement with the simulations in Fig.~\ref{fig:occ_f>0}, right, where we plot single-fermion correlations $\langle c^{\dag}_{j}c_{j+r}\rangle$ for $\tau = \pm 4$. In both cases the decay is exponential, but for $\tau =-4$ it is significantly slower.

\subsection{Entanglement entropy and central charge}

As the low energy excitations of the fermionic and of the bosonic fluids in the 2F model are expected to display Luttinger liquid behaviour, it is natural to monitor the behaviour of the entanglement entropy $S(\ell) = -\text{tr}[\rho_\ell \log \rho_\ell]$ of the first $\ell$ sites of the system as a function of $\ell$ while varying the control parameter $\tau$, as mentioned in the main text. The reason for it is the relation between the entanglement entropy of the first $l$ sites of a system in OBC and whose low energy behaviour is captured by a conformal field theory, and the central charge of the conformal field theory itself, summarized in the formula: 
\begin{equation}
\label{eq:entropy}
S(\ell)=\frac{c}{6}\log\left[\frac{2L}{\pi}\sin\left(\frac{\pi \ell}{L}\right)\right]+A + C_f\ev*{c^{\dag}_{\ell+1}c^{\dag}_{\ell}+\text{h.c.}}+
C_b \ev*{c^{\dag}_{\ell+2}c^{\dag}_{\ell +1}c_{\ell}c_{\ell-1}+\text{h.c.}},
\end{equation} 
where $c$ is the central charge, $L$ is the system size and $A$ and $C_{f,b}$ are constants.

\begin{figure}[t]
\centering
\includegraphics[width=0.75\columnwidth,clip]{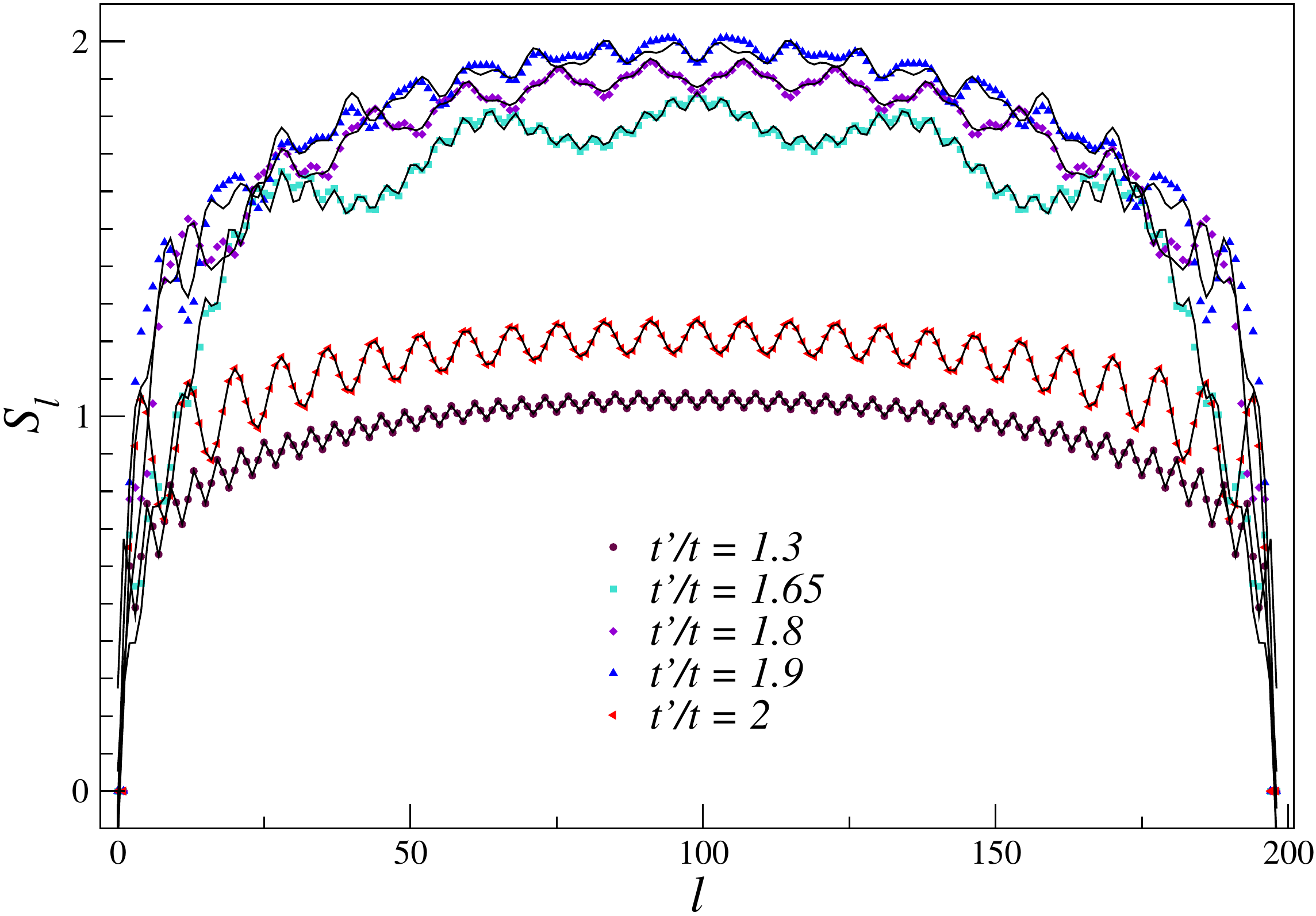}
\caption{Fits of the entropy : symbols are DMRG data, black lines are fits with Eq.~\eqref{eq:entropy}.}
\label{fig:entropy_fit}
\end{figure}

Some example fits of the entanglement entropy profile in the three identified phases are provided in Fig. $\ref{fig:entropy_fit}$; their outcome agrees with the low energy theories predicted for the three liquid phases by the 2F model: while the $F$ and $P_\pi$ phases display only two gapless points and are thus described by a single $c=1$ Luttinger liquid theory, the entanglement entropy profiles in the $C$ phase are compatible with a $c=2$ phase, consistently with the additional gapless mode arising from the coexistence of gapless single-particle and pair excitations in the system.

For $t'/t = 1.65$, one also nicely sees the typical real space picture valid both for local density and local kinetic energies. One has large bumps corresponding to pairs and small bumps corresponding to unpaired fermions.

\end{document}